\documentclass[a4paper,fleqn]{article}
\usepackage{graphics}
\newcommand{\be}{\begin{equation}}
\newcommand{\ee}{\end{equation}}
\newcommand{\ba}[1]{\begin{array}{#1}}
\newcommand{\ea}{\end{array}}
\newcommand{\n}[1]{\overline{#1}}
\newcommand{\bra}[1][\Psi_0]{\ensuremath{\langle #1 |}}
\newcommand{\ket}[1][\Psi_0]{\ensuremath{| #1 \rangle}}
\newcommand{\braket}[2]{\ensuremath{\langle #1 | #2 \rangle}}
\newcommand{\proj}[1]{\ensuremath{|#1\rangle\langle #1|}}
\newcommand{\neel}{N\'{e}el}

\newcommand{\mas}{\renewcommand{\arraystretch}{1.3}}

\newcommand{\zii}{\ensuremath{Z^{\makebox[0em][l]{$\rightarrow$}\raisebox{0.8ex}{$\rightarrow$}}}}
\newcommand{\zoo}{\ensuremath{Z^{\makebox[0em][l]{$\leftarrow $}\raisebox{0.8ex}{$\leftarrow $}}}}
\newcommand{\zio}{\ensuremath{Z^{\makebox[0em][l]{$\leftarrow $}\raisebox{0.8ex}{$\rightarrow$}}}}
\newcommand{\zoi}{\ensuremath{Z^{\makebox[0em][l]{$\rightarrow$}\raisebox{0.8ex}{$\leftarrow $}}}}
\renewcommand{\phi}{\varphi}
\newcommand{\verthaoii}{        %
\begin{minipage}{18mm}          %
\unitlength1mm                  %
\begin{picture}(18,16)          %
\thinlines                      %
\put( 0, 8){\line(1, 0){10}}    %
\put(10, 8){\line(1, 1){7}}     %
\put(10, 8){\line(1,-1){7}}     %
\put(10, 8){\circle*{3}}        %
\put( 3, 8){\line(2, 1){4}}     %
\put( 3, 8){\line(2,-1){4}}     %
\put(12,10){\line(1, 2){2}}     %
\put(12,10){\line(2, 1){4}}     %
\put(12, 6){\line(1,-2){2}}     %
\put(12, 6){\line(2,-1){4}}     %
\end{picture}                   %
\end{minipage}}                 %

\newcommand{\verthaiii}{        %
\begin{minipage}{18mm}          %
\unitlength1mm                  %
\begin{picture}(18,16)          %
\thinlines                      %
\put( 0, 8){\line(1, 0){10}}    %
\put(10, 8){\line(1, 1){7}}     %
\put(10, 8){\line(1,-1){7}}     %
\put(10, 8){\circle*{3}}        %
\put( 7, 8){\line(-2, 1){4}}    %
\put( 7, 8){\line(-2,-1){4}}    %
\put(12,10){\line( 1, 2){2}}    %
\put(12,10){\line( 2, 1){4}}    %
\put(12, 6){\line( 1,-2){2}}    %
\put(12, 6){\line( 2,-1){4}}    %
\end{picture}                   %
\end{minipage}}                 %

\newcommand{\verthaoio}{        %
\begin{minipage}{18mm}          %
\unitlength1mm                  %
\begin{picture}(18,16)          %
\thinlines                      %
\put( 0, 8){\line(1, 0){10}}    %
\put(10, 8){\line(1, 1){7}}     %
\put(10, 8){\line(1,-1){7}}     %
\put(10, 8){\circle*{3}}        %
\put( 3, 8){\line( 2, 1){4}}    %
\put( 3, 8){\line( 2,-1){4}}    %
\put(12,10){\line( 1, 2){2}}    %
\put(12,10){\line( 2, 1){4}}    %
\put(16, 2){\line(-1, 2){2}}    %
\put(16, 2){\line(-2, 1){4}}    %
\end{picture}                   %
\end{minipage}}                 %

\newcommand{\verthaiio}{        %
\begin{minipage}{18mm}          %
\unitlength1mm                  %
\begin{picture}(18,16)          %
\thinlines                      %
\put( 0, 8){\line(1, 0){10}}    %
\put(10, 8){\line(1, 1){7}}     %
\put(10, 8){\line(1,-1){7}}     %
\put(10, 8){\circle*{3}}        %
\put( 7, 8){\line(-2, 1){4}}    %
\put( 7, 8){\line(-2,-1){4}}    %
\put(12,10){\line( 1, 2){2}}    %
\put(12,10){\line( 2, 1){4}}    %
\put(16, 2){\line(-1, 2){2}}    %
\put(16, 2){\line(-2, 1){4}}    %
\end{picture}                   %
\end{minipage}}                 %

\newcommand{\verthaooi}{        %
\begin{minipage}{18mm}          %
\unitlength1mm                  %
\begin{picture}(18,16)          %
\thinlines                      %
\put( 0, 8){\line(1, 0){10}}    %
\put(10, 8){\line(1, 1){7}}     %
\put(10, 8){\line(1,-1){7}}     %
\put(10, 8){\circle*{3}}        %
\put( 3, 8){\line( 2, 1){4}}    %
\put( 3, 8){\line( 2,-1){4}}    %
\put(16,14){\line(-1,-2){2}}    %
\put(16,14){\line(-2,-1){4}}    %
\put(12, 6){\line( 1,-2){2}}    %
\put(12, 6){\line( 2,-1){4}}    %
\end{picture}                   %
\end{minipage}}                 %

\newcommand{\verthaioi}{        %
\begin{minipage}{18mm}          %
\unitlength1mm                  %
\begin{picture}(18,16)          %
\thinlines                      %
\put( 0, 8){\line(1, 0){10}}    %
\put(10, 8){\line(1, 1){7}}     %
\put(10, 8){\line(1,-1){7}}     %
\put(10, 8){\circle*{3}}        %
\put( 7, 8){\line(-2, 1){4}}    %
\put( 7, 8){\line(-2,-1){4}}    %
\put(16,14){\line(-1,-2){2}}    %
\put(16,14){\line(-2,-1){4}}    %
\put(12, 6){\line( 1,-2){2}}    %
\put(12, 6){\line( 2,-1){4}}    %
\end{picture}                   %
\end{minipage}}                 %

\newcommand{\verthaooo}{        %
\begin{minipage}{18mm}          %
\unitlength1mm                  %
\begin{picture}(18,16)          %
\thinlines                      %
\put( 0, 8){\line(1, 0){10}}    %
\put(10, 8){\line(1, 1){7}}     %
\put(10, 8){\line(1,-1){7}}     %
\put(10, 8){\circle*{3}}        %
\put( 3, 8){\line( 2, 1){4}}    %
\put( 3, 8){\line( 2,-1){4}}    %
\put(16,14){\line(-1,-2){2}}    %
\put(16,14){\line(-2,-1){4}}    %
\put(16, 2){\line(-1, 2){2}}    %
\put(16, 2){\line(-2, 1){4}}    %
\end{picture}                   %
\end{minipage}}                 %

\newcommand{\verthaioo}{        %
\begin{minipage}{18mm}          %
\unitlength1mm                  %
\begin{picture}(18,16)          %
\thinlines                      %
\put( 0, 8){\line(1, 0){10}}    %
\put(10, 8){\line(1, 1){7}}     %
\put(10, 8){\line(1,-1){7}}     %
\put(10, 8){\circle*{3}}        %
\put( 7, 8){\line(-2, 1){4}}    %
\put( 7, 8){\line(-2,-1){4}}    %
\put(16,14){\line(-1,-2){2}}    %
\put(16,14){\line(-2,-1){4}}    %
\put(16, 2){\line(-1, 2){2}}    %
\put(16, 2){\line(-2, 1){4}}    %
\end{picture}                   %
\end{minipage}}                 %

\newcommand{\verthcoii}{        %
\begin{minipage}{18mm}          %
\unitlength1mm                  %
\begin{picture}(18,16)          %
\thinlines                      %
\put( 0, 8){\line(1, 0){10}}    %
\put(10, 8){\line(1, 1){7}}     %
\put(10, 8){\line(1,-1){7}}     %
\put(10, 8){\circle{3}}         %
\put( 3, 8){\line(2, 1){4}}     %
\put( 3, 8){\line(2,-1){4}}     %
\put(12,10){\line(1, 2){2}}     %
\put(12,10){\line(2, 1){4}}     %
\put(12, 6){\line(1,-2){2}}     %
\put(12, 6){\line(2,-1){4}}     %
\end{picture}                   %
\end{minipage}}                 %

\newcommand{\verthciii}{        %
\begin{minipage}{18mm}          %
\unitlength1mm                  %
\begin{picture}(18,16)          %
\thinlines                      %
\put( 0, 8){\line(1, 0){10}}    %
\put(10, 8){\line(1, 1){7}}     %
\put(10, 8){\line(1,-1){7}}     %
\put(10, 8){\circle{3}}         %
\put( 7, 8){\line(-2, 1){4}}    %
\put( 7, 8){\line(-2,-1){4}}    %
\put(12,10){\line( 1, 2){2}}    %
\put(12,10){\line( 2, 1){4}}    %
\put(12, 6){\line( 1,-2){2}}    %
\put(12, 6){\line( 2,-1){4}}    %
\end{picture}                   %
\end{minipage}}                 %

\newcommand{\verthcoio}{        %
\begin{minipage}{18mm}          %
\unitlength1mm                  %
\begin{picture}(18,16)          %
\thinlines                      %
\put( 0, 8){\line(1, 0){10}}    %
\put(10, 8){\line(1, 1){7}}     %
\put(10, 8){\line(1,-1){7}}     %
\put(10, 8){\circle{3}}         %
\put( 3, 8){\line( 2, 1){4}}    %
\put( 3, 8){\line( 2,-1){4}}    %
\put(12,10){\line( 1, 2){2}}    %
\put(12,10){\line( 2, 1){4}}    %
\put(16, 2){\line(-1, 2){2}}    %
\put(16, 2){\line(-2, 1){4}}    %
\end{picture}                   %
\end{minipage}}                 %

\newcommand{\verthciio}{        %
\begin{minipage}{18mm}          %
\unitlength1mm                  %
\begin{picture}(18,16)          %
\thinlines                      %
\put( 0, 8){\line(1, 0){10}}    %
\put(10, 8){\line(1, 1){7}}     %
\put(10, 8){\line(1,-1){7}}     %
\put(10, 8){\circle{3}}         %
\put( 7, 8){\line(-2, 1){4}}    %
\put( 7, 8){\line(-2,-1){4}}    %
\put(12,10){\line( 1, 2){2}}    %
\put(12,10){\line( 2, 1){4}}    %
\put(16, 2){\line(-1, 2){2}}    %
\put(16, 2){\line(-2, 1){4}}    %
\end{picture}                   %
\end{minipage}}                 %

\newcommand{\verthcooi}{        %
\begin{minipage}{18mm}          %
\unitlength1mm                  %
\begin{picture}(18,16)          %
\thinlines                      %
\put( 0, 8){\line(1, 0){10}}    %
\put(10, 8){\line(1, 1){7}}     %
\put(10, 8){\line(1,-1){7}}     %
\put(10, 8){\circle{3}}         %
\put( 3, 8){\line( 2, 1){4}}    %
\put( 3, 8){\line( 2,-1){4}}    %
\put(16,14){\line(-1,-2){2}}    %
\put(16,14){\line(-2,-1){4}}    %
\put(12, 6){\line( 1,-2){2}}    %
\put(12, 6){\line( 2,-1){4}}    %
\end{picture}                   %
\end{minipage}}                 %

\newcommand{\verthcioi}{        %
\begin{minipage}{18mm}          %
\unitlength1mm                  %
\begin{picture}(18,16)          %
\thinlines                      %
\put( 0, 8){\line(1, 0){10}}    %
\put(10, 8){\line(1, 1){7}}     %
\put(10, 8){\line(1,-1){7}}     %
\put(10, 8){\circle{3}}         %
\put( 7, 8){\line(-2, 1){4}}    %
\put( 7, 8){\line(-2,-1){4}}    %
\put(16,14){\line(-1,-2){2}}    %
\put(16,14){\line(-2,-1){4}}    %
\put(12, 6){\line( 1,-2){2}}    %
\put(12, 6){\line( 2,-1){4}}    %
\end{picture}                   %
\end{minipage}}                 %

\newcommand{\verthcooo}{        %
\begin{minipage}{18mm}          %
\unitlength1mm                  %
\begin{picture}(18,16)          %
\thinlines                      %
\put( 0, 8){\line(1, 0){10}}    %
\put(10, 8){\line(1, 1){7}}     %
\put(10, 8){\line(1,-1){7}}     %
\put(10, 8){\circle{3}}         %
\put( 3, 8){\line( 2, 1){4}}    %
\put( 3, 8){\line( 2,-1){4}}    %
\put(16,14){\line(-1,-2){2}}    %
\put(16,14){\line(-2,-1){4}}    %
\put(16, 2){\line(-1, 2){2}}    %
\put(16, 2){\line(-2, 1){4}}    %
\end{picture}                   %
\end{minipage}}                 %

\newcommand{\verthcioo}{        %
\begin{minipage}{18mm}          %
\unitlength1mm                  %
\begin{picture}(18,16)          %
\thinlines                      %
\put( 0, 8){\line(1, 0){10}}    %
\put(10, 8){\line(1, 1){7}}     %
\put(10, 8){\line(1,-1){7}}     %
\put(10, 8){\circle{3}}         %
\put( 7, 8){\line(-2, 1){4}}    %
\put( 7, 8){\line(-2,-1){4}}    %
\put(16,14){\line(-1,-2){2}}    %
\put(16,14){\line(-2,-1){4}}    %
\put(16, 2){\line(-1, 2){2}}    %
\put(16, 2){\line(-2, 1){4}}    %
\end{picture}                   %
\end{minipage}}                 %

\newcommand{\verthprodaiio}{    %
\begin{minipage}{18mm}          %
\unitlength1mm                  %
\begin{picture}(18,16)          %
\thinlines                      %
\put( 8, 8){\line( 1, 0){10}}   %
\put( 8, 8){\line(-1, 1){7}}    %
\put( 8, 8){\line(-1,-1){7}}    %
\put( 8, 8){\circle*{3}}        %
\put( 2, 2){\line( 2, 1){4}}    %
\put( 2, 2){\line( 1, 2){2}}    %
\put( 6,10){\line(-1, 2){2}}    %
\put( 6,10){\line(-2, 1){4}}    %
\put(11, 8){\line( 2, 1){4}}    %
\put(11, 8){\line( 2,-1){4}}    %
\end{picture}                   %
\end{minipage}}                 %

\newcommand{\verthprodaoio}{    %
\begin{minipage}{18mm}          %
\unitlength1mm                  %
\begin{picture}(18,16)          %
\thinlines                      %
\put( 8, 8){\line( 1, 0){10}}   %
\put( 8, 8){\line(-1, 1){7}}    %
\put( 8, 8){\line(-1,-1){7}}    %
\put( 8, 8){\circle*{3}}        %
\put( 2, 2){\line( 2, 1){4}}    %
\put( 2, 2){\line( 1, 2){2}}    %
\put( 6,10){\line(-1, 2){2}}    %
\put( 6,10){\line(-2, 1){4}}    %
\put(15, 8){\line(-2, 1){4}}    %
\put(15, 8){\line(-2,-1){4}}    %
\end{picture}                   %
\end{minipage}}                 %

\newcommand{\verthprodbooi}{    %
\begin{minipage}{18mm}          %
\unitlength1mm                  %
\begin{picture}(18,16)          %
\thinlines                      %
\put( 0, 8){\line(1, 0){10}}    %
\put(10, 8){\line(1, 1){7}}     %
\put(10, 8){\line(1,-1){7}}     %
\put(10, 8){\circle{3}}         %
\put( 3, 8){\line( 2, 1){4}}    %
\put( 3, 8){\line( 2,-1){4}}    %
\put(16,14){\line(-1,-2){2}}    %
\put(16,14){\line(-2,-1){4}}    %
\put(12, 6){\line( 1,-2){2}}    %
\put(12, 6){\line( 2,-1){4}}    %
\end{picture}                   %
\end{minipage}}                 %

\newcommand{\verthprodbioi}{    %
\begin{minipage}{18mm}          %
\unitlength1mm                  %
\begin{picture}(18,16)          %
\thinlines                      %
\put( 0, 8){\line(1, 0){10}}    %
\put(10, 8){\line(1, 1){7}}     %
\put(10, 8){\line(1,-1){7}}     %
\put(10, 8){\circle{3}}         %
\put( 7, 8){\line(-2, 1){4}}    %
\put( 7, 8){\line(-2,-1){4}}    %
\put(16,14){\line(-1,-2){2}}    %
\put(16,14){\line(-2,-1){4}}    %
\put(12, 6){\line( 1,-2){2}}    %
\put(12, 6){\line( 2,-1){4}}    %
\end{picture}                   %
\end{minipage}}                 %

\newcommand{\verthprod}{        %
\begin{minipage}{26mm}          %
\unitlength1mm                  %
\begin{picture}(26,16)          %
\thinlines                      %
\put( 8, 8){\line( 1, 0){10}}   %
\put( 8, 8){\line(-1, 1){7}}    %
\put( 8, 8){\line(-1,-1){7}}    %
\put( 8, 8){\circle*{3}}        %
\put( 2, 2){\line( 2, 1){4}}    %
\put( 2, 2){\line( 1, 2){2}}    %
\put( 6,10){\line(-1, 2){2}}    %
\put( 6,10){\line(-2, 1){4}}    %
\put(18, 8){\line(1, 1){7}}     %
\put(18, 8){\line(1,-1){7}}     %
\put(18, 8){\circle{3}}         %
\put(24,14){\line(-1,-2){2}}    %
\put(24,14){\line(-2,-1){4}}    %
\put(20, 6){\line( 1,-2){2}}    %
\put(20, 6){\line( 2,-1){4}}    %
\end{picture}                   %
\end{minipage}}                 %

\title{Spin-$\frac{3}{2}$ models on the Cayley tree --
       optimum ground state approach}
\author{H.~Niggemann \and J.~Zittartz}
\date{\small Institut f\"ur Theoretische Physik, Z\"ulpicher Str. 77,
             D-50937 K\"oln}
\begin{document}
\maketitle
\begin{abstract}
We present a class of {\em optimum ground states} for
spin-$\frac{3}{2}$ models on the Cayley tree with coordination
number 3. The interaction is restricted to nearest neighbours and
contains 5 continuous parameters. For all values of these parameters
the Hamiltonian has parity invariance, spin-flip invariance, and
rotational symmetry in the $xy$-plane of spin space. The global ground
states are constructed in terms of a 1-parametric
{\em vertex state model}, which is a direct generalization of the
well-known {\em matrix product ground state} approach. By using
recursion relations and the transfer matrix technique we derive exact
analytical expressions for local fluctuations and longitudinal and
transversal two-point correlation functions.
\end{abstract}

\renewcommand{\thefootnote}{}
\footnotetext{Work performed within the research program of the
Sonderforschungsbereich 341, K\"{o}ln-Aachen-J\"{u}lich}
\renewcommand{\thefootnote}{\arabic{footnote}}

\section{Introduction}
\begin{figure}[t]
\begin{center}
\resizebox{8cm}{!}{\includegraphics{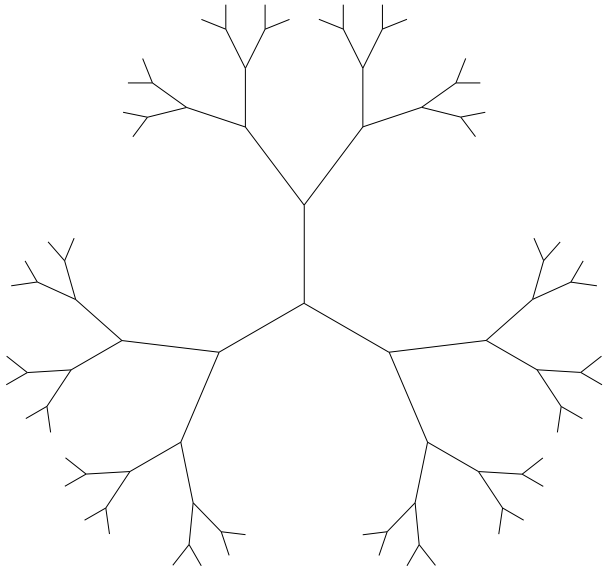}}
\end{center}
\caption{A finite Cayley tree with coordination number~3.}
\label{fcayley}
\end{figure}

The Cayley tree belongs to the category of {\em pseudo-lattices}
\cite{moraal}. Unlike regular lattices, which are usually defined in
terms of periodic structures, the Cayley tree is generated by the
following {\em recursive} scheme:
\begin{enumerate}
\parskip0ex
\item A {\em Cayley branch} of order $1$ is a single lattice site.
\item A {\em Cayley branch} of order $n$ is defined as a lattice site
with $K\!+\!1$ bonds, to which $K$ Cayley branches of order $n\!-\!1$
are attached, i.e.\ the branch has 1 unconnected bond.
\item A {\em Cayley tree} of order $n$ is given by a {\em central}
lattice site with $K\!+\!1$ bonds, to which $K\!+\!1$ branches of
order $n\!-\!1$ are attached.
\end{enumerate}
$K$ is called the {\em connectivity}, $K\!+\!1$ is the
{\em coordination number} of the Cayley tree. Figure~\ref{fcayley}
shows a finite Cayley tree with coordination number~3. In the
thermodynamic limit $n\to\infty$ the Cayley tree is also known as the
{\em Bethe lattice}.

An important property of the Cayley tree is that there is exactly one
path from a lattice site $i$ to another lattice site $j$. The
{\em distance} $|i\!-\!j|$ between $i$ and $j$ is simply the number of
edges on this path. We can use the distance function to divide the
Cayley tree into two disjoint sublattices. Denote the central site by
$i_0$. An arbitrary lattice site $i$ belongs to the sublattice
${\cal L}_A$ if $|i_0\!-\!i|$ is even, otherwise $i$ belongs to the
sublattice ${\cal L}_B$. It is easy to see that every site
$i\in {\cal L}_A$ has only nearest neighbours in ${\cal L}_B$ and vice
versa, so $({\cal L}_A,{\cal L}_B)$ is a {\em bipartite decomposition}
of the Cayley tree. Note that it is valid for any connectivity $K$.

Due to the hierarchical structure of the lattice, the partition
function of a classical statistical model on the Cayley tree can
usually be calculated by using recursion relations, provided the local
interaction has finite range and the number of states at each lattice
site is also finite. This is a substantial step beyond models on the
chain, as it allows to construct exactly solvable models 
{\em with arbitrary coordination number}. The most important drawback
of the Cayley tree is that for large system sizes $\frac{1}{K}$th of
the lattice sites are boundary sites\footnote{For small system sizes
the percentage of boundary sites is even larger.}. As a consequence
the physics is heavily influenced by boundary effects. There is no
canonical way to impose periodic boundary conditions.

In this work we investigate the ground state problem of a class of
quantum spin-$\frac{3}{2}$ models on the Cayley tree with
connectivity $K=2$. The paper is organized as follows. 
Section~\ref{definition} contains the definition of the Hamiltonian
and discusses its parameters and symmetries. The global ground state
is constructed explicitly in section~\ref{construction} in terms of a
{\em vertex state model}. Vertex state models are graphical
realizations of so-called {\em optimum ground states}, which
simultaneously minimize all local interaction operators. Ground state
properties, i.e.\ single-spin and two-point expectation values are
presented in section~\ref{properties}. As shown in appendix~A the
calculation of ground state expectation values leads to a classical
vertex model on the Cayley tree, which can be solved exactly. Finally
we summarize our results in section~\ref{summary}.

\section{Model definition}
\label{definition}
The model is defined on the Cayley tree with coordination number
$3$. A spin-$\frac{3}{2}$ is located at each lattice site. These
spin variables are coupled by nearest neighbour interaction terms
$h_{ij}$, which are all equal, they only act on different pairs of
lattice sites. The global Hamiltonian $H$ is the sum of all these
local interactions, so the system is completely homogeneous.

The local interaction is the same as in our previous works on the
hexagonal lattice \cite{hexag32} and the two-leg ladder
\cite{laddr32}. Hence we shall be very brief here. For the
construction of {\em optimum ground states} it is advantageous to
write the interaction operator in terms of projectors onto its
eigenstates
\be
\label{hij}
\mas
\ba{lcl}
h_{ij}& = &
  \lambda_3 \,\left(\, \proj{v_3} + \proj{v_{-3}} \,\right)\, + \\
&&\lambda_2^{-\sigma} \,\left(\, \proj{v_2^{-\sigma}} +
                                 \proj{v_{-2}^{-\sigma}} \,\right)\, + \\
&&\lambda_{12}^+ \,\left(\, \proj{v_{12}^+} + \proj{v_{-12}^+} \,\right)\, + \\
&&\lambda_{02}^{-\sigma} \, \proj{v_{02}^{-\sigma}} \, .
\ea\ee
If we use the following notation for the canonical basis states of a
single spin-$\frac{3}{2}$,
\be
\mas
\ba{lcl@{\hspace{2cm}}lcl}
S^z \ket[3]     &=&  \frac{3}{2} \ket[3] &
S^z \ket[\n{3}] &=& -\frac{3}{2} \ket[\n{3}] \\
S^z \ket[1]     &=&  \frac{1}{2} \ket[1] &
S^z \ket[\n{1}] &=& -\frac{1}{2} \ket[\n{1}] \, ,
\ea
\ee
the eigenstates used in (\ref{hij}) are given by
\be
\label{lexcited}
\mas
\ba{lcl}
\ket[v_3] &=& \ket[33] \\
\ket[v_{-3}] &=& \ket[\n{33}] \\
\ket[v_2^{-\sigma}] &=& \ket[31]-\sigma\ket[13] \\
\ket[v_{-2}^{-\sigma}] &=& \ket[\n{31}]-\sigma\ket[\n{13}] \\
\ket[v_{12}^+] &=& a \ket[11]
    - \,\left(\, \ket[3\n{1}]+\ket[\n{1}3] \,\right)\, \\
\ket[v_{-12}^+] &=& a \ket[\n{11}]
    - \,\left(\, \ket[\n{3}1]+\ket[1\n{3}] \,\right)\, \\
\ket[v_{02}^{-\sigma}] &=& \sigma a^2 
    \,\left(\, \ket[1\n{1}]-\sigma\ket[\n{1}1] \,\right)\, -
    \,\left(\, \ket[3\n{3}]-\sigma\ket[\n{3}3] \,\right)\, .
\ea
\ee
The parameters
$\lambda_3,\lambda_2^{-\sigma},\lambda_{12}^+,\lambda_{02}^{-\sigma}$
are real and positive and the {\em superposition parameter} $a$ is
real. $\sigma$ is a discrete parameter, which can only take the values
$\pm 1$. Thus the total number of continuous parameters is 5, which
includes a trivial scale, so there are 4 non-trivial interaction
parameters.

For all values of the parameters $h_{ij}$ (\ref{hij})
commutes with the pair magnetization operator $S^z_i+S^z_j$ and with
the parity operator $P_{ij}$, which interchanges the spins at sites
$i$ and $j$. Therefore the local interaction (\ref{hij}) has
rotational symmetry in the $xy$-plane of spin space and is parity
invariant. In addition, corresponding eigenstates with magnetization
$m$ and $-m$ carry the same $\lambda$-coefficient, so $h_{ij}$ is also
invariant under a spin-flip $S^z\to -S^z$. In particular, no external
magnetic field is applied.

As all $\lambda$-parameters are positive, (\ref{hij}) is a positive
semi-definite operator, i.e.\ all its eigenvalues are non-negative.
The two-spin states (\ref{lexcited}) are the excited local eigenstates
of $h_{ij}$, the remaining 9 eigenstates are local ground states,
i.e.\ the corresponding eigenvalue is zero. Since the Hamiltonian $H$
is the sum of positive semi-definite operators, the global ground
state energy $E_0$ is non-negative, too. In the next section we shall
show that $E_0$ is in fact zero and the corresponding global ground
state will be constructed explicitly.

At the {\em isotropic point}, $a=-\sqrt{3}$ and $\sigma=-1$, the
$\lambda$-parameters can be adjusted so that $h_{ij}$ has the form
\be
h_{ij} =             {\bf S}_i \cdot {\bf S}_j 
  + \frac{116}{243}( {\bf S}_i \cdot {\bf S}_j )^2
  +  \frac{16}{243}( {\bf S}_i \cdot {\bf S}_j )^3
  +  \frac{55}{108} \, .
\ee
Obviously, this operator has complete $SO(3)$ symmetry. It simply
projects onto all states with $({\bf S}_i+{\bf S}_j)^2=3(3+1)$.
This case has already been investigated in \cite{aklt}. Its ground
state is known as the {\em valence bond solid} (VBS) ground state.
As shown in \cite{aklt}, it has exponentially decaying correlation
functions, no \neel\ order, and there is an energy gap between the
ground state and the lowest excitations. This is consistent with our
results presented in Section~\ref{properties}.

\section{Construction of the global ground state}
\label{construction}
In this section we construct the exact ground state of the present
model. It is an {\em optimum ground state} \cite{ksz}--\cite{laddr32},
i.e.\ it is not only the ground state of the global Hamiltonian $H$,
but also of every local interaction operator $h_{ij}$. For spin chains
such global states can be generated by using so-called
{\em matrix product ground states} (MPG) \cite{ksz,chain32}. A
generalization of the MPG concept to arbitrary lattices is given by
{\em vertex state models} \cite{chain32}--\cite{laddr32}, which have
been used to construct optimum ground states on the hexagonal lattice
and on the two-leg ladder.

In order to construct the global ground state for the present model,
we assign the following set of vertices to each site on the first
sublattice of the Cayley tree:
\be
\label{vert1}
\renewcommand{\arraystretch}{4}
\ba{rcr@{\hspace{2cm}}rcr}
\verthaooo&:& \sigma a \ket[3]     &
\verthaiii&:& \sigma a \ket[\n{3}] \\
\verthaooi&:&          \ket[1]     &
\verthaiio&:&          \ket[\n{1}] \\
\verthaioo&:&          \ket[1]     &
\verthaoii&:&          \ket[\n{1}] \\
\verthaoio&:&          \ket[1]     &
\verthaioi&:&          \ket[\n{1}] \makebox[0em][l]{\, .}
\ea\ee
The corresponding vertices on the second sublattice are
\be
\label{vert2}
\renewcommand{\arraystretch}{4}
\ba{rcr@{\hspace{2cm}}rcr}
\verthcooo&:&       a \ket[3]     &
\verthciii&:&\sigma a \ket[\n{3}] \\
\verthcooi&:&         \ket[1]     &
\verthciio&:&\sigma   \ket[\n{1}] \\
\verthcioo&:&         \ket[1]     &
\verthcoii&:&\sigma   \ket[\n{1}] \\
\verthcoio&:&         \ket[1]     &
\verthcioi&:&\sigma   \ket[\n{1}] \makebox[0em][l]{\, .}
\ea\ee
Unlike a {\em classical} vertex model, each vertex has a single-spin
state $\alpha\ket[m]$ as its value (or `weight'), where
\be
m=\frac{1}{2}(\# \mbox{ of outgoing arrows} -
              \# \mbox{ of incoming arrows}) \, .
\ee
The parameters $a$ and $\sigma$ are the same as in (\ref{lexcited}).
Both sets of vertices differ only with respect to the positions of the
$\sigma$-coefficients. Note that (\ref{vert1}) and (\ref{vert2}) are
the same as on the hexagonal lattice \cite{hexag32}, only the global
lattice topology is different.

The global ground state $\ket$ is generated by concatenating the
vertices at all lattice sites. As in usual classical vertex models of
statistical physics, the connecting bond between adjacent lattice
sites is summed out. The generic product of vertex weights is replaced
by the tensorial product in spin space:
\be
\ba{lcl}
\verthprod
& = & \verthprodaiio \otimes \verthprodbooi \\
& + & \verthprodaoio \otimes \verthprodbioi
\ea
\ee
It can be shown that the resulting global state is indeed an optimum
ground state of $H$ by collecting all two-spin states which are
generated by all possible concatenations of neighbouring
vertices\footnote{Common prefactors have been omitted.}:
\be
\label{lground}
\mas
\ba{l@{\hspace{2cm}}l}
\ket[31] + \sigma \ket[13] &
\ket[\n{31}] + \sigma \ket[\n{13}] \\
\ket[11] + a \ket[3\n{1}] &
\ket[\n{11}] + a \ket[\n{3}1] \\
\ket[11] + a \ket[\n{1}3] &
\ket[\n{11}] + a \ket[1\n{3}] \\
\ket[1\n{1}] + \sigma a^2 \ket[3\n{3}] &
\ket[\n{1}1] + \sigma a^2 \ket[\n{3}3] \\
\ket[1\n{1}] + \sigma \ket[\n{1}1] \, . & 
\ea
\ee
These 9 two-spin states are orthogonal on all excited local states
(\ref{lexcited}), so (\ref{lground}) are the local ground states of
$h_{ij}$. Therefore it is clear that any projection of $\ket$ onto the
space of two adjacent lattice sites is a linear combination of local
ground states. This yields
\be
h_{ij} \ket = 0
\ee
for all nearest neighbours $i$ and $j$ and hence also
\be
H \ket = 0 \, .
\ee
Since zero is a lower bound of the global ground state energy, the
constructed vertex state model is indeed an optimum ground state of
the global Hamiltonian $H$.

In contrast to regular lattices there is no canonical way to impose
periodic boundary conditions on the Cayley tree, so open boundary
conditions are used. In this case, the vertices on the boundary sites
('leafs') emanate {\em external bonds} which are not summed
out. Independent of the arrow configuration on these external bonds,
the resulting vertex state model is always an optimum ground state of
$H$. Thus the ground state degeneracy grows exponentially with system
size.

\section{Properties of the ground state}
\label{properties}
Each arrow configuration $\lbrace b \rbrace$ on the external bonds
generates an optimum ground state $\ket[\Psi_0^{\lbrace b \rbrace}]$
of $H$. The calculation of ground state expectation values requires
taking the average over all these configurations
\be
\label{average}
\langle A \rangle_{\Psi_0} = \frac{1}{B} \; \sum_{\lbrace b \rbrace} \;
\langle A \rangle_{\Psi_0^{\lbrace b \rbrace}} \, .
\ee
$B$ denotes the total number of such configurations. By using the
techniques developed in appendices~A and~B, ground state expectation
values of arbitrary single-spin observables and two-spin correlation
functions can be obtained exactly. As described in appendix~A, the
average over all boundary arrow configurations is performed
automatically. Note that the formulae given below hold for {\em all}
system sizes, on which the considered observables can be applied
(cf.\ appendix~B). In particular the results are valid in the
thermodynamic limit. This pathological effect occurs only on the
Cayley tree, optimum ground states on regular lattices exhibit a
non-trivial finite-size behaviour.

The first interesting expectation values are the components of the
canonical spin operator
\be
\label{vectorexpect}
\langle S_i^x \rangle_{\Psi_0}=
\langle S_i^y \rangle_{\Psi_0}=
\langle S_i^z \rangle_{\Psi_0}= 0 \, .
\ee
This is the expected result as the boundary conditions and the global
ground states preserve the symmetries of the Hamiltonian, in
particular spin-flip symmetry and rotational invariance in the
$xy$-plane of spin space. (\ref{vectorexpect}) implies that the total
magnetization is zero, so the global ground state is antiferromagnetic.

Although the local magnetization vanishes, its fluctuations are
non-trivial
\be
\langle \left( S_i^z \right)^2 \rangle_{\Psi_0} =
\frac{9}{4} - \frac{6}{3+a^2} \, ,
\ee
which increases monotonically from $\frac{1}{4}$ to $\frac{9}{4}$ as a
function of $a^2$, thus covering the full range of possible values.
Because of rotational symmetry in the $xy$-plane of spin space we also
obtain
\be
\langle \left( S_i^x \right)^2 \rangle_{\Psi_0} =
\langle \left( S_i^y \right)^2 \rangle_{\Psi_0} =
\frac{1}{2}\left[ \frac{3}{2}\left(\frac{3}{2}+1\right)-
\langle \left( S_i^z \right)^2 \rangle_{\Psi_0} \right]  =
\frac{3}{4}+\frac{3}{3+a^2} \, .
\ee

\begin{figure}[t]
\begin{center}
\tabcolsep0ex
\begin{tabular}{rc}
\raisebox{2.5cm}{$\ba{ll} \xi_{\rm l}^{-1} \\ \xi_{\rm t}^{-1} \ea$} &
\resizebox{8cm}{!}{\includegraphics{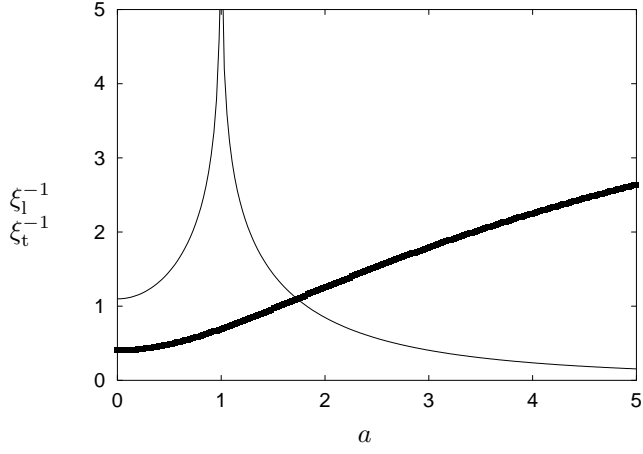}} \\
& $a$
\end{tabular}
\end{center}
\caption{Inverse longitudinal (thin) and transversal (thick)
correlation length as a function of the parameter $a$.}
\label{fplotxi}
\end{figure}

In appendix~B the transfer matrix technique has been employed to
compute two-point correlation functions. The result for the
longitudinal correlation function is
\be
\label{szszcorrgen}
\langle S_i^z S_j^z \rangle_{\Psi_0} =
-\left(\frac{1+3a^2}{6+2a^2}\right)^2 \cdot
 \left(\frac{1- a^2}{3+ a^2}\right)^{|i-j|-1} \, ,
\ee
where $|i\!-\!j|$ is the distance between lattice sites $i$ and
$j$. Note that the correlation decays exponentially as a function of
the distance and its sign alternates if $a^2>1$.
The corresponding longitudinal correlation length can be read off
from equation~(\ref{szszcorrgen}). Its inverse is given by
\be
\label{xil}
\xi_{\rm l}^{-1} = \ln \left|\frac{3+a^2}{1-a^2}\right| \, ,
\ee
which is plotted as a function of the parameter $a$ in
figure~\ref{fplotxi}. Note the divergence at $a=1$. In this special
case, longitudinal correlations between spins with a distance of $2$
or larger are completely absent (cf.~equation~(\ref{szszcorrgen})).

As the longitudinal one, the transversal two-spin correlation function
decays exponentially as a function of the distance:
\be
\langle S_i^x S_j^x \rangle_{\Psi_0} =
\left(\frac{2+\sqrt{3}\,\sigma a}{3+a^2}\right)^2 \cdot
\left(\frac{2}{3+a^2}\right)^{|i-j|-1} \, .
\ee
Thus the inverse transversal correlation length is
\be
\label{xit}
\xi_{\rm t}^{-1} = \ln \frac{3+a^2}{2} \, .
\ee
It is also plotted in figure~\ref{fplotxi}, together with the
longitudinal one. Both inverse correlation length, $\xi_{\rm l}^{-1}$
and $\xi_{\rm t}^{-1}$, are non-zero for all finite values of $a$,
hence the model is never critical.

\begin{figure}[t]
\begin{center}
\tabcolsep0ex
\begin{tabular}{rc}
\raisebox{2.5cm}{$\langle S_i^z S_{i+1}^z \rangle_{\Psi_0}$} &
\resizebox{8cm}{!}{\includegraphics{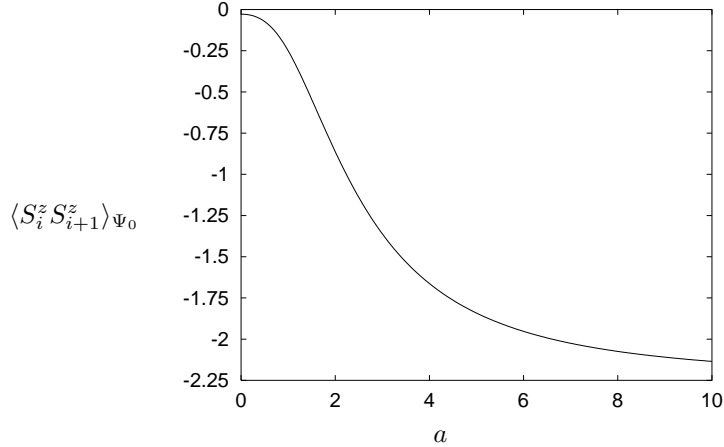}} \\
& $a$
\end{tabular}
\end{center}
\caption{Longitudinal nearest-neighbour correlation as a function 
of the parameter $a$.}
\label{fplotnnc}
\end{figure}

In the special case $|i\!-\!j|=1$ equation (\ref{szszcorrgen}) yields
the longitudinal nearest-neighbour correlation
$\langle S_i^z S_{i+1}^z \rangle_{\Psi_0}$.
Starting at $-\frac{1}{36}$ for $a=0$ it decreases monotonically and
approaches $-\frac{9}{4}$ asymptotically for large values of $a$, as
shown in figure~\ref{fplotnnc}.

There are three noteworthy special points in the parameter space. The
first one is $a^2=1$. In this case all non-vanishing vertices of the
classical vertex model which corresponds to the inner product
$\braket{\Psi_0}{\Psi_0}$ (cf.\ appendix~A) have the same weight,
namely 1. This corresponds to infinitely high temperature in the
language of classical vertex models, i.e.\ disorder is maximal. The
vanishing of longitudinal correlations for $a^2=1$
(cf.\ figure~\ref{fplotxi}) is consistent with this interpretation.

The next interesting special case is the isotropic point
$a=-\sqrt{3},\,\sigma=-1$ where we can adjust the $\lambda$-parameters
so that the local interaction operator (\ref{hij}) has full
$SO(3)$ symmetry. As mentioned in section~\ref{definition}, this model
has already been investigated in \cite{aklt}. The reported inverse
correlation length $\xi_{\rm iso}^{-1} = \ln 3$ for open boundary
conditions coincides with the results obtained from (\ref{xil}) and
(\ref{xit}).

Finally we consider the limit $a^2\to\infty$. As can be seen from
(\ref{vert1}) and (\ref{vert2}) the global ground state is dominated
by only four vertices in this limit, namely
\[
\renewcommand{\arraystretch}{2}
\ba{rcr@{\hspace{2cm}}rcr}
\verthaooo&:& \sigma a \ket[3]     &
\verthaiii&:& \sigma a \ket[\n{3}] \\
\multicolumn{6}{l}{\hspace{-\mathindent}\mbox{on the first sublattice and}} \\
\verthcooo&:&        a \ket[3]     &
\verthciii&:& \sigma a \ket[\n{3}]
\ea
\]
on the second one. The two ways, in which these vertices can be
combined on the Cayley tree, represent the two different \neel\
states. All spins on the first sublattice are in the $\ket[3]$ state,
the others are in the $\ket[\n{3}]$ state, and vice versa. Therefore
this special case is called the {\em \neel\ limit}. Note that the
ground state degeneracy is higher than in the generic case. In
addition to the degeneracy due to the open boundary conditions there
is also a `bulk degeneracy', as some of the local ground states
(\ref{lground}) become simple tensor products of single-spin states.

\section{Summary}
\label{summary}
We have investigated the ground state problem of a class of
antiferromagnetic spin-$\frac{3}{2}$ models on the Cayley tree with
coordination number~3. Apart from the lattice topology the Hamiltonian
is the same as in our previous works on the hexagonal lattice
\cite{hexag32} and the two-leg ladder \cite{laddr32}. It is defined in
terms of the nearest neighbour interaction, which contains 5
continuous parameters and has parity invariance, spin-flip invariance,
and rotational invariance in the $xy$-plane of spin space.

Due to the open boundary conditions the ground state degeneracy grows
exponentially with system size. We have constructed the global ground
states explicitly by using the {\em vertex state model} approach. These
are so-called {\em optimum ground states}, i.e.\ they are not only
ground states of the global Hamiltonian, but simultaneously minimize
all local interaction operators. The vertex state model contains a
continuous parameter $a$, which controls $z$-axis anisotropy, and a
discrete parameter $\sigma=\pm 1$.

The calculation of ground state expectation values leads to a
classical vertex model on the same lattice as the original quantum
spin model. Due to the hierarchical structure of the Cayley tree this
classical model can be solved exactly by using recursion relations and
the transfer matrix technique. The result of our calculations is that
the model has vanishing sublattice magnetization and exponentially
decaying correlation functions. Exact formulae for the nearest
neighbour correlation, the longitudinal and transversal correlation
lengths, and for the fluctuations of the magnetization have been
derived. For special values of $a$ and $\sigma$ the global ground
state coincides with the so-called {\em valence bond solid} (VBS)
ground state.

\appendix
\renewcommand{\theequation}{\Alph{section}.\arabic{equation}}
\setcounter{equation}{0}
\renewcommand{\thesection}{Appendix \Alph{section}:}

\section{Calculation of single-spin expectation values}
For the moment we consider only one of the many global
ground states, namely the vertex state model with all boundary arrows
pointing {\em out of} the leaf sites. Denote this ground state by
$\ket[\Psi_0^*]$. Within this ground state, the expectation value of
an observable $A_i$, which only acts on the spin at lattice site $i$,
is defined as
\be
\label{generalexpect}
\langle A_i \rangle_{\Psi_0^*}=
\frac{\bra[\Psi_0^*] A_i \ket[\Psi_0^*]}{\braket{\Psi_0^*}{\Psi_0^*}}
\, .
\ee
The denominator can be interpreted as two identical vertex state
models on top of each other, representing the bra- and the ket-vector,
respectively. Since the vertices at each lattice site generate only
local single-spin states, the inner product can be taken separately at
each lattice site before the interior bonds are summed out. Hence
$\braket{\Psi_0^*}{\Psi_0^*}$ can be interpreted as the partition
function of a {\em classical} vertex model with vertices defined as
\be
\label{classical}
\ba{c}
\resizebox{8cm}{!}{\includegraphics{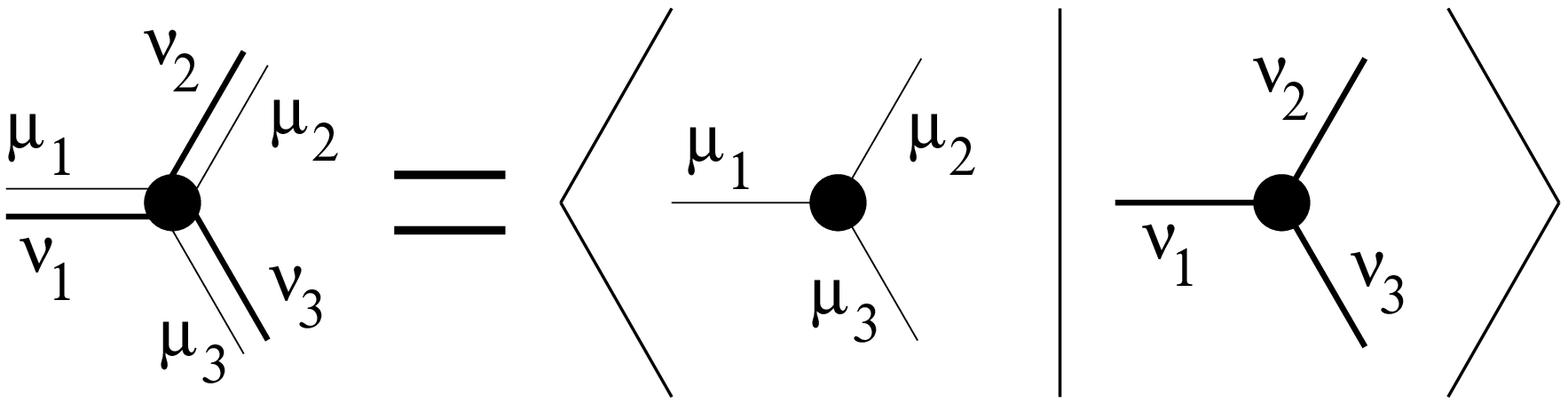}}
\ea
\, .
\ee
These vertices have the following properties:
\begin{itemize}
\parskip0ex
\item The vertex weights are real numbers, not single-spin states.
\item There are two arrow variables on each bond, originating from the
bra- and the ket-vector.
\item The vertices are identical on both sublattices of the Cayley
tree as $\sigma^2=1$.
\item Only 20 of the 64 different vertices have a non-vanishing
weight since the inner product between different $S^z$-eigenstates is
zero.
\end{itemize}

\begin{figure}[t]
\begin{center}
\resizebox{5cm}{!}{\includegraphics{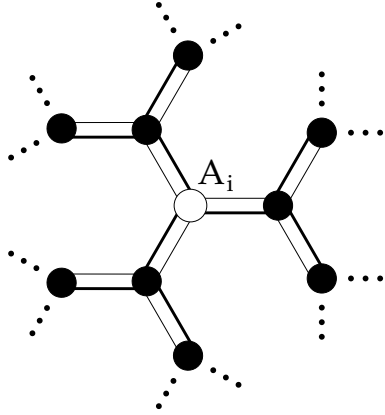}}
\end{center}
\caption{Graphical representation of single-spin expectation values.}
\label{fsingleexpect}
\end{figure}

The numerator of (\ref{generalexpect}) corresponds to the same
classical vertex model as the denominator, except for site $i$ where
the vertices are modified. At this special lattice site, the classical
vertices are given by inserting the operator $A_i$ between the bra-
and the ket-vector on the r.h.s.\ of (\ref{classical}). The graphical
representation of $\bra[\Psi_0^*] A_i \ket[\Psi_0^*]$ is shown in
figure~\ref{fsingleexpect}. The affected site $i$ is surrounded by
three unmodified {\em branches} of the classical vertex model. In the
following, these branches are denoted by $\zoo_n,\zii_n,\zio_n$, and
$\zoi_n$, defined as
\be
\label{branchdef}
Z_n^{\mu\nu} = \ba{c}
\resizebox{!}{3cm}{\includegraphics{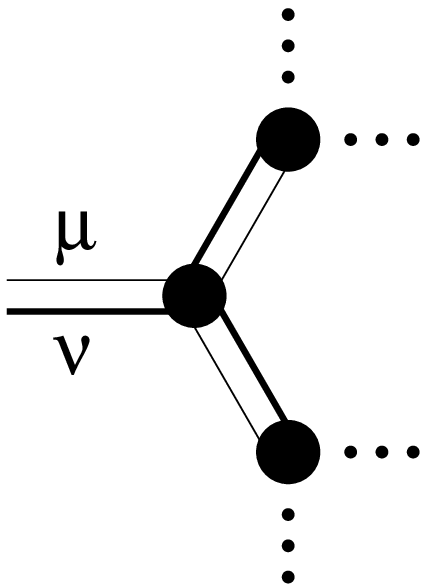}}
\ea
\ee
Here the lower index $n$ denotes the order, i.e.\ the branch contains
$2^n\!-\!1$ lattice sites. Due to the hierarchical structure the
values of (\ref{branchdef}) can be calculated by using recursion
relations. Each $Z_{n+1}$ is given by concatenating two copies of
$Z_n$ to a single classical vertex and summing out the connecting
bonds. This yields the recursion relations
\be
\label{recursion}
\ba{lcl}
\zoo_{n+1}=\left(\zoo_n+\zii_n\right)^2+(a^2-1)\left(\zii_n\right)^2
           +2\zio_n\zoi_n \\
\zii_{n+1}=\left(\zoo_n+\zii_n\right)^2+(a^2-1)\left(\zoo_n\right)^2
           +2\zio_n\zoi_n \\
\zio_{n+1}=\left(\zoo_n+\zii_n\right)\zio_n \\
\zoi_{n+1}=\left(\zoo_n+\zii_n\right)\zoi_n \, .
\ea
\ee
In order to actually solve these coupled equations it is necessary to
specify the initial values $\zoo_0,\zii_0,\zio_0,\zoi_0$ which are of
course determined by the boundary conditions. For $\ket[\Psi_0^*]$,
which is defined as having all boundary arrows pointing out of the
leaf sites, the correct choice would be to set $\zii_0=1$ and the
other three initial values to zero\footnote{%
Multiplying the set of initial values with a common non-zero constant
leaves all expectation values unchanged. So $\zii_0=1$ is a convenient
normalization.}.

However, in order to calculate ground state expectation values,
definition~(\ref{average}) requires the summation over all arrow
configurations on the external bonds. It turns out that this summation
can be carried out {\em automatically} by using the initial values
\begin{eqnarray}
\label{boundary0a}
&& \zoo_0=\zii_0=1 \qquad\mbox{and} \\
\label{boundary0b}
&& \zoi_0=\zio_0=0 \, .
\end{eqnarray}
(\ref{boundary0a}) ensures that configurations with incoming and
outgoing arrows are weighted equally and (\ref{boundary0b}) eliminates
all `mixed' terms, i.e.\ terms where bra- and ket-vector are different.

As a consequence of these initial values we obtain
\begin{eqnarray}
\label{boundaryna}
&& \zoo_n=\zii_n \qquad\mbox{and} \\
\label{boundarynb}
&& \zoi_n=\zio_n=0
\end{eqnarray}
for all $n\geq 0$. This is immediately clear from the recursion
relations. Inserting (\ref{boundaryna}),(\ref{boundarynb}) into
(\ref{recursion}) yields
\be
\zoo_{n+1}=\zii_{n+1}=(3+a^2) \left( \zoo_n \right)^2 \, ,
\ee
which has the solution
\be
\label{zsolution}
\zoo_n=\zii_n=(3+a^2)^{(2^n-1)} \, .
\ee
Note that the exponent $2^n\!-\!1$ is simply the number of lattice
sites of the branch. It is now straightforward to calculate the ground
state expectation value of the local observable $A_i$ as
\be
\label{singleexpectfinal}
\langle A_i \rangle_{\Psi_0} = \frac{Z(A_i)}{Z} \, .
\ee
$Z(A_i)$ is given by attaching the solution (\ref{zsolution}) to all
bonds of the classical vertex modified by the operator $A_i$
(cf.\ figure~\ref{fsingleexpect}). The denominator is simply
\be
Z=Z(1)=2 (3+a^2)^N \, ,
\ee
$N$ being the total number of lattice sites. Due to the product
structure of the solution (\ref{zsolution}) the results for all
expectation values are {\em independent} of $N$. Additional powers of
$3+a^2$ drop out in the quotient (\ref{singleexpectfinal}). As shown
in appendix~B, a similar effect also occurs in the general case of
$k$-point correlation functions. This means that there are
{\em no finite size effects}. Note that although the classical vertex
model has this simple product solution, the underlying vertex state
model generates a highly non-trivial global ground state.

Equation (\ref{boundarynb}) means that the classical vertex model
contains {\em no unequal arrow pairs} on its bonds. The reason for
this exact vanishing is that only classical vertices with zero or two
unequal arrow pairs have a non-vanishing weight. Two of the three
bonds of a leaf site are external bonds, which do not carry unequal
arrow pairs (due to the open boundary conditions). So each leaf site
provides the next hierarchy of vertices only with equal arrow pairs,
and so on. Thus unequal arrow pairs can never enter the classical
vertex model. This is in contrast to the hexagonal lattice
\cite{hexag32}, where unequal arrow pairs are in general present,
albeit in a very low concentration for most values of the parameter
$a$. This is a {\em dynamical} effect.

\section{Calculation of two-spin expectation values}
In this appendix we extend the technique developed in appendix~A to
two-point correlation functions $\langle A_i B_j \rangle_{\Psi_0}$.
$A_i$ and $B_j$ are observables which only act on lattice sites $i$
and $j$, respectively.

\begin{figure}[t]
\begin{center}
\resizebox{8cm}{!}{\includegraphics{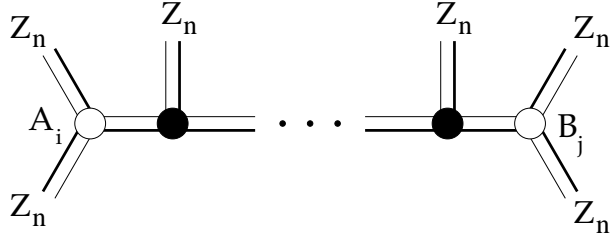}}
\end{center}
\caption{Graphical representation of two-spin correlation functions.}
\label{ftwoexpect}
\end{figure}

The Cayley tree is free of loops, so there is exactly one path from
site $i$ to site $j$. Figure~\ref{ftwoexpect} shows the topological
structure of the corresponding classical vertex model. The branches
$Z_n^{\mu\nu}$ are known from (\ref{boundarynb}) and (\ref{zsolution}),
so the remaining problem is the summation along the path from $i$ to
$j$. If $|i\!-\!j|$ denotes the distance between these two lattice sites
the path is given by the $(|i\!-\!j|\!-\!1)$-fold product of the
periodicity element\footnote{%
Strictly speaking the order $n$ of the branch $Z_n^{\mu\nu}$ can be
different for each vertex along the path from site $i$ to site
$j$. However, for the same reason as in appendix~A the order of these
branches is completely irrelevant for the calculation of expectation
values.}
\be
\label{periodicity}
\ba{c}
\resizebox{2cm}{!}{\includegraphics{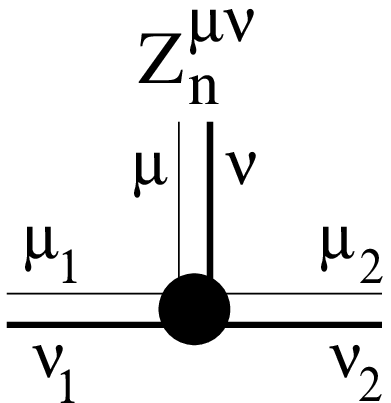}}
\ea
\ee
The bonds $\mu_1\nu_1$ and $\mu_2\nu_2$ connect this element to the
previous and to the next periodicity element on the path. The idea is
to interpret (\ref{periodicity}) as a {\em transfer matrix} $T$
\cite{ksz,chain32,laddr32} and to use its eigenbasis to calculate
$T^{|i-j|-1}$.

Each bond carries two binary arrow variables, so $T$ is a $4\times
4$-matrix. If the following mapping of arrow configurations to matrix
indices is used
\be
\ba{lcl@{\hspace{2cm}}lcl}
\makebox[0em][l]{\raisebox{-0.4ex}{$\leftarrow $}}
                 \raisebox{ 0.4ex}{$\leftarrow $}&:&1 &
\makebox[0em][l]{\raisebox{-0.4ex}{$\rightarrow$}}
                 \raisebox{ 0.4ex}{$\leftarrow $}&:&3 \\
\makebox[0em][l]{\raisebox{-0.4ex}{$\rightarrow$}}
                 \raisebox{ 0.4ex}{$\rightarrow$}&:&2 &
\makebox[0em][l]{\raisebox{-0.4ex}{$\leftarrow $}}
                 \raisebox{ 0.4ex}{$\rightarrow$}&:&4
\ea
\ee
then the transfer matrix is given by
\be
T=c_Z \left( \ba{cccc}
            2 & 1+a^2 & 0 & 0 \\
            1+a^2 & 2 & 0 & 0 \\
            0     & 0 & 2 & 0 \\
            0     & 0 & 0 & 2 \ea \right) \, .
\ee
$c_Z=(3+a^2)^{(2^n-1)}$ is the prefactor due to the attached branch
$Z_n^{\mu\nu}$. The eigenvalues $\chi_k$ and the corresponding
normalized eigenvectors $\ket[u_k]$ of this real symmetric matrix are
\be
\label{teigen}
\ba{rcl@{\hspace{2cm}}rcl}
\chi_1&=&(3+a^2)c_Z &
\ket[u_1]&=&(1,\phantom{-}1,0,0)/\sqrt{2} \\
\chi_2&=&(1-a^2)c_Z &
\ket[u_2]&=&(1,-1,0,0)/\sqrt{2} \\
\chi_3&=& 2c_Z     &
\ket[u_3]&=&(0,0,1,0) \\
\chi_4&=& 2c_Z     &
\ket[u_4]&=&(0,0,0,1) \, .
\ea
\ee
This eigensystem can now be used to compute the desired powers of the
transfer matrix as
\be
\label{tpower}
T^{|i-j|-1}=\sum_k \; \chi_k^{|i-j|-1} \; \ket[u_k]\bra[u_k] \, .
\ee
The final step is to assemble all components of the classical vertex
model as shown in figure~\ref{ftwoexpect}. To the left/right-hand side
of (\ref{tpower}) we attach the vertices modified by the observable
$A_i$/$B_j$, respectively. Unmodified branches $Z_n$ are attached to
the remaining four bonds. In analogy to (\ref{singleexpectfinal}) the
desired expectation value is given by
\be
\label{twoexpectfinal}
\langle A_i B_j \rangle_{\Psi_0} = \frac{Z(A_i,B_j)}{Z} \, .
\ee
The numerator is the partition function of the modified vertex model
calculated in this appendix and the denominator is the partition
function of the {\em un}modified vertex model $Z=2 (3+a^2)^N$. As in
the previous appendix, the average over all configurations of the
boundary arrows is performed automatically.

Note that the order of the branch $Z_n$ in (\ref{periodicity}) enters
the calculation only via the factor $c_Z$ which appears in all
eigenvalues of $T$ (\ref{teigen}). However, $c_Z$ drops out in the
quotient (\ref{twoexpectfinal}), so the order of the branches
attached to the path from site $i$ to site $j$ is irrelevant. The
same holds for the four branches on the l.h.s.\ and on the r.h.s.\ of
figure~\ref{ftwoexpect}. Therefore the expectation value
(\ref{twoexpectfinal}) only depends on the observables $A_i$ and $B_j$
and on their distance $|i\!-\!j|$. The size of the rest of the Cayley
tree has no influence.

The technique developed in this appendix can be easily generalized to
$k$-point correlation functions. If $k$ is finite there is always a
decomposition of the full lattice into a finite number of
\begin{itemize}
\parskip0ex
\item paths,
\item modified vertices, and
\item unmodified branches.
\end{itemize}
These objects can be dealt with separately. The resulting expectation
values depend only on the observables themselves and the length of the
paths between the vertices modified by these observables. The size of
the unmodified branches is irrelevant. In this sense there are no
finite size effects.

\end{document}